\documentclass[conference]{IEEEtran}
\IEEEoverridecommandlockouts
\usepackage{xcolor}  
\usepackage{tabularx}
\usepackage{tikz}
\usepackage{amsmath}
\usepackage{graphicx}
\usepackage{booktabs}
\usepackage{latexsym} 
\usepackage{array}    
\usepackage{multirow}
\usepackage{algorithm}
\usepackage[noend]{algpseudocode}
\usepackage{threeparttable}
\usepackage{paralist}   
\usepackage{xspace}
\usepackage{color}
\usepackage{adjustbox}
\usepackage{wasysym}
\usepackage{rotating}   
\usepackage{listings}
\usepackage{tabu}
\usepackage{amssymb}
\usepackage{amsfonts}
\usepackage{textcomp}
\usepackage{pifont}
\usepackage{framed}
\usepackage[shortlabels]{enumitem}
\usepackage{makecell}
\usepackage{soul}
\setenumerate[1]{itemsep=0pt,partopsep=0pt,parsep=\parskip,topsep=2pt}
\setitemize[1]{itemsep=0pt,partopsep=0pt,parsep=\parskip,topsep=2pt}
\setdescription{itemsep=0pt,partopsep=0pt,parsep=\parskip,topsep=2pt}

\usepackage{url}            
\usepackage{hyperref}
\usepackage{flushend}

\usepackage{xpatch}
\usepackage{graphics}
\usepackage[utf8]{inputenc}
\usepackage{cleveref}
\usepackage{circledsteps}
\crefname{section}{§}{§§}
\Crefname{section}{§}{§§}
\newtheorem{definition}{Definition}
\makeatletter
\chardef\TPT@@@asteriskcatcode=\catcode`*
\catcode`*=11
\xpatchcmd{\threeparttable}
  {\TPT@hookin{tabular}}
  {\TPT@hookin{tabular}\TPT@hookin{tabu}}
  {}{}
\catcode`*=\TPT@@@asteriskcatcode
\makeatother

\usepackage{tcolorbox}
\tcbuselibrary{breakable, skins}
\tcbset{%
  label begin/.style={label={#1}},
  label end/.style={after upper=\label{#1}}
}
\newtcolorbox[%
auto counter]{mybox}[2][]{%
  enhanced jigsaw,
  breakable,
  #1}

\newcommand{\distance}{2pt}
\newif\ifANNOYMIZE
\ANNOYMIZEtrue

\newif\ifACM
\ACMtrue  

\newif\ifUSENIX
\USENIXtrue

\ifACM
\newcommand{\myfig}{Fig.\xspace}
\else
\newcommand{\myfig}{Fig.\xspace}
\fi

\newcommand{\mytab}{Table\xspace}

\ifACM
\newcommand{\mysec}{\S}
\else
\newcommand{\mysec}{Section\xspace}
\fi

\ifACM

\else

\fi

\ifACM
\newcommand{\mydef}{Definition\xspace}
\else
\newcommand{\mydef}{Definition\xspace}
\fi

\newcommand{\fixme}[1]{{\color{black}{#1}}}

\definecolor{cadmiumgreen}{rgb}{0.0, 0.42, 0.24}

\newcommand{\name}{DeFiScope\xspace}
\newcommand{\tool}{\name}
\newsavebox{\bigimage} 

\newcommand{\cmark}{\ding{51}\xspace}%
\newcommand{\xmark}{\ding{55}\xspace}%

\ifCLASSOPTIONcompsoc
 \usepackage[caption=false,font=normalsize,labelfont=sf,textfont=sf]{subfig}
\else
 \usepackage[caption=false,font=footnotesize]{subfig}
\fi
\def\BibTeX{{\rm B\kern-.05em{\sc i\kern-.025em b}\kern-.08em
    T\kern-.1667em\lower.7ex\hbox{E}\kern-.125emX}}

\begin{document}
%
\title{\tool: Detecting Various DeFi Price Manipulations with LLM Reasoning}
\title{Detecting Various DeFi Price Manipulations with LLM Reasoning}

\author{
\IEEEauthorblockN{Juantao Zhong\IEEEauthorrefmark{1}, Daoyuan Wu\IEEEauthorrefmark{1}$^1$\thanks{1: Daoyuan Wu is the co-first author.}, Ye Liu\IEEEauthorrefmark{2}$^2$\thanks{2: Ye Liu is the corresponding author.}, Maoyi Xie\IEEEauthorrefmark{3}, Yang Liu\IEEEauthorrefmark{3}, Yi Li\IEEEauthorrefmark{3}, and Ning Liu\IEEEauthorrefmark{4}}
\IEEEauthorblockA{\IEEEauthorrefmark{1} Lingnan University, \texttt{\{ericzhong|daoyuanwu\}@ln.edu.hk} \\
\IEEEauthorrefmark{2} Singapore Management University, \texttt{yeliu@smu.edu.sg}\\
\IEEEauthorrefmark{3} Nanyang Technological University, \texttt{maoyi001@e.ntu.edu.sg, \{yangliu|yi\_li\}@ntu.edu.sg}\\
\IEEEauthorrefmark{4} City University of Hong Kong, City University of Hong Kong Shenzhen Research Institute, \texttt{ninliu@cityu.edu.hk}\\
}
}

\maketitle

\begin{abstract}

DeFi (Decentralized Finance) is one of the most important applications of today's cryptocurrencies and smart contracts.
It manages hundreds of billions in Total Value Locked (TVL) on-chain, yet it remains susceptible to common DeFi price manipulation attacks.
Despite state-of-the-art (SOTA) systems like DeFiRanger and DeFort, we found that they are less effective to non-standard price models in custom DeFi protocols, which account for 44.2\% of the 95 DeFi price manipulation attacks reported over the past three years.

In this paper, we introduce the first LLM-based approach, \tool, for detecting DeFi price manipulation attacks in both standard 
and custom price models.
Our insight is that large language models (LLMs) have certain intelligence to abstract price calculation from
\fixme{smart contract source code}
and infer the trend of token price changes based on the extracted price models.
To further strengthen LLMs in this aspect, we leverage Foundry to synthesize on-chain data and use it to fine-tune a DeFi price-specific LLM. 
Together with the high-level DeFi operations recovered from low-level transaction data, \tool detects various DeFi price manipulations according to systematically mined patterns.
\fixme{Experimental results show that \tool achieves a high recall of 80\% on real-world attacks, a precision of 96\% on suspicious transactions, and zero false alarms on benign transactions, significantly outperforming SOTA approaches.}
Moreover, we evaluate \tool's cost-effectiveness and demonstrate its practicality by helping our industry partner confirm 147 real-world price manipulation attacks, including discovering 81 previously unknown historical incidents.


\end{abstract}

\section{Introduction}
\label{sec:introduction}


\fixme{DeFi represents a form of finance that eliminates traditional intermediaries by utilizing smart contracts on a blockchain. A smart contract~\cite{zou2019smart} is a self-executing program with the terms of the agreement directly written into code, deployed on a distributed, decentralized blockchain network~\cite{dinh2018untangling}.
These contracts power decentralized applications such as Automated Market Makers (AMMs), where token prices are determined by on-chain pricing mechanisms
rather than centralized order books.
While price manipulation is a pervasive risk in traditional finance, DeFi pricing mechanisms create more attack surfaces: adversaries can strategically add or remove liquidity, or impact token supply by minting or burning, to exploit vulnerabilities in pricing mechanisms, ultimately distorting prices of cryptocurrencies~\cite{cointelegraph}.}
To detect price manipulation attacks in DeFi applications, researchers have proposed several state-of-the-art (SOTA) approaches, primarily focusing on transaction monitoring-based methods that offer real-time protection, e.g., DeFiRanger~\cite{DeFiRanger23} and DeFort~\cite{DeFort24}.

However, our analysis indicates that these SOTA approaches are less effective to non-standard price models in custom DeFi protocols.
This is because they typically require token exchange rates to identify abnormal price changes, which are only suitable to calculate under standard price models such as CPMM (Constant Product Market Makers) and Stableswap Invariant (detailed in \mysec\ref{sec:backg_priceModel}).
Unfortunately, our subsequent evaluation shows that 44.2\% of the 95 DeFi price manipulation attacks reported in the past three years used non-standard price models.
Hence, \textit{instead of explicitly calculating the exchange rate for a pair of tokens, we aim to capture only the abnormal price fluctuations of tokens}, which can be directly derived from the high-level price model and changes in token balances.

To achieve this objective, we sought help from large language models (LLMs), considering that their trained intelligence might aid in inferring price changes associated with DeFi operations during the transaction process.
As such, we introduce \tool,
\fixme{a source code-level detector and}
the first LLM-based approach to detecting DeFi price manipulation attacks in both standard and custom price models.
\tool features several novel designs, such as constructing a transfer graph to recover high-level DeFi operations (\mysec\ref{sec:defiOperation}) and systematically mining price manipulation patterns across 
four prevalent types
of DeFi applications (\mysec\ref{sec:finalDetect}).
Among them, the key design is to fine-tune a DeFi price-specific LLM (\mysec\ref{sec:priceInfer}), where we propose (i) simulating transactions using Foundry~\cite{Foundry} to generate our own fine-tuning data, and (ii) conducting a Chain-of-Thought (CoT)-style fine-tuning that integrates both on-chain data and the price context.
The ablation study shows that fine-tuning increases the overall detection effectiveness by up to 31\% compared to the baseline LLM (under \tool) without fine-tuning.

To thoroughly evaluate \tool's effectiveness and practicality, we conduct both a benchmark and a large-scale experiment.
In the benchmark experiment, we collected 95 real-world price manipulation attacks from multiple sources and compared \tool with three SOTA tools.
The results show that \tool achieves a superior detection accuracy (recall) of 80\%, significantly higher than 51.6\% with DeFiRanger, 52.6\% with DeFort, and 35.8\% with DeFiTainter~\cite{DeFiTainter23} (a static source code analysis tool).
\fixme{In our large-scale experiment, we evaluated \tool on two datasets provided by our industry partner and DeFort~\cite{DeFort24}. The first dataset consists of 968 suspicious transactions, covering a variety of DeFi attacks beyond price manipulations. On this dataset, \tool flagged 153 potential price manipulation attacks, 147 of which were manually verified as true positives, yielding a precision of 96\%.}
\fixme{Notably, 81 of these attacks are previously undocumented price manipulation incidents.}
\fixme{The second dataset involves 96,800 benign transactions, on which \tool produced zero false alarms. Regarding efficiency, \tool incurs an average time overhead of 2.5 seconds per transaction across both suspicious and benign transactions, with a per-request LLM inference cost of only \$0.0107.}

In sum, this paper makes the following contributions:
\begin{itemize}[leftmargin=*,noitemsep,topsep=0pt]
\item We introduce the first LLM-based approach, \tool, for automatic on-chain price manipulation attack detection.
In particular, LLMs, when properly fine-tuned, have the capability to abstract price models and infer price changes. 

\item To support \tool's detection, we propose a graph-based method to recover high-level DeFi operations and systematically mine eight price manipulation patterns.

\item We extensively evaluate \tool with three real-world transaction datasets, showing \tool's superior performance over SOTA tools in terms of precision and recall. 
\end{itemize}

\noindent
\textbf{Artifact.}
To facilitate open science and future research, 
all source code, experimental results, and supplementary material are available in
\url{https://github.com/AIS2Lab/DeFiScope}.

\section{Background and Motivation}
\label{sec:background}


On blockchains, two primary account types exist, i.e., externally owned accounts (EOAs), controlled by individuals using private keys, and contract accounts (CAs), governed by their contract code. 
External transactions are initiated externally by EOAs, while internal transactions occurs when a smart contract calls another contract internally. 
Smart contracts exchange information with other smart contracts through internal transactions, where
a transaction involving communication between contracts can be regarded as a sequence of function calls, utilized by DeFi protocols for interoperability.
For simplicity, we denote as \emph{user invocation} an internal transaction between the user-controlled smart contracts and other smart contracts, which plays a key role in price manipulation attacks.

\subsection{Price Models}
\label{sec:backg_priceModel}

A price model represents the pricing mechanism within the DeFi application, which are typically expressed as equations correlating the price of a certain token with the balances of and the total supply of various tokens as well as other constant. 


\noindent\textbf{Constant Product Market Maker (CPMM).}
CPMM is one of the most prevalent DeFi AMMs and used in many well-known DEXs~\cite{Uniswap,PancakeSwap}. It maintains liquidity through a constant function
\fixme{$x*y=k$,}
where $k$ is a constant, and
\fixme{$x$ and $y$}
represent the reserves of two distinct assets $token_x$ and $token_y$ in a liquidity pool. In a swap operation, let $\Delta x$ amount of $token_x$ can exchange for $\Delta y$ amount of $token_y$, resulting in $(x+\Delta x)*(y-\Delta y)=k$ where the instantaneous price of $token_x$ denominated in $token_y$ is $P_{x,y} = \frac{y}{x}$~\cite{DeFiRanger23,gogol2024sok}.

\noindent\textbf{Stableswap Invariant.}
The Stableswap Invariant~\cite{StableSwap}, widely used in Curve AMMs~\cite{Curve}, and is defined as follows:
\begin{align}
\label{Stableswap}
    \frac{a\prod_{i=1}^{n} x_i}{(D/n)^{n}} \cdot D^{n-1} \cdot \sum_{i=1}^{n} x_i + \prod_{i=1}^{n} x_i &= \frac{a\prod_{i=1}^{n} x_i}{(D/n)^n} \cdot D^n + \frac{D^n}{n^n} \nonumber
\end{align}
where $a$ indicates a constant amplification coefficient and $n$ is the number of tokens in the liquidity pool, while $D$ represents 
the total amount of tokens in the pool when token prices are equal, and $x_i$ denotes the current reserve of $token_i$. 
\fixme{The curve is nearly flat in the middle part, while price changes at the extremes resemble CPMM behavior.}

\noindent\textbf{Custom Price Model.}
Besides the mentioned
common price models, DeFi protocols 
can customize their own pricing mechanisms, which are often more diverse and complicated. 
For example, 
UwULend~\cite{uwulend_official} offers lending service, permitting users to borrow sUSDe by depositing cryptocurrencies as collateral. 
To ensure the user can repay,
examining whether the total value of sUSDe are below that of the collateral when lending it is necessary.
In this process, UwULend calculates the price of sUSDe by the following custom price model, distinct from the aforementioned CPMM and Stableswap Invariant models:
\begin{equation}
\label{PriceOfsUSDe}
\begin{split}
    P_{sUSDe} = median(\{IP_{USDe,Pool_1},\cdots, IP_{USDe,Pool_5}, \\EMAP_{USDe,Pool_1},\cdots, EMAP_{USDe,Pool_5}\})
\end{split}
\end{equation}
where $P_{sUSDe}$ is the price of sUSDe, calculated as 
the median of a set of prices relevant to USDe in five liquidity pools\footnote{$Pool_{FRAXUSDe}$~\cite{FRAXUSDe}, $Pool_{USDeUSDC}$~\cite{USDeUSDC}, $Pool_{USDeDAI}$~\cite{USDeDAI}, $Pool_{USDecrvUSD}$~\cite{USDecrvUSD}, and $Pool_{GHOUSDe}$~\cite{GHOUSDe}.}.
$EMAP_{USDe,Pool_i}$ is the Exponential Moving Average (EMA)~\cite{EMA} price of USDe in the i-th liquidity pool but remains a constant value within a transaction block.
In contrast, $IP_{USDe,Pool_i}$ refers to the instantaneous price of USDe in the i-th liquidity pool. 
Unlike the EMA, the latter price is more volatile and vulnerable to fluctuations in tokens' balance. 

\subsection{A Motivating Example}
\label{sec:motivatingexample}


Our approach is motivated by a real-world price manipulation attack on UwULend~\cite{MotivationAnalysis,MotivationTrax} in 2024.
The root cause of the attack is the flawed price dependency in \cref{PriceOfsUSDe} related to two functions \texttt{borrow} and \texttt{liquidationCall}.
By swapping a large amount of USDe into five liquidity pools, the attacker deflated the instant price of USDe while the EMA prices kept unchanged, resulting in a lower median price for sUSDe.
Hence, the attack was able to borrow an exceedingly large amount of sUSDe using the same amount of collateral. 
Then, swapping back could increase the calculated price of sUSDe, resulting in the attacker's collateral being unable to repay the debt, thereby allowing liquidation and acquiring deposited collateral with a bonus, leading to a \$19M loss.

While detecting a price manipulation attack, existing tools generally define the exchange rate between two different tokens as their price.
For example, the price of sUSDe could be calculated by dividing the amount of WETH deposited into the protocol
by the amount of sUSDe borrowed from it.
Yet, defining abnormal price change range or even precisely capturing token prices is not a trivial task.

\begin{figure*}[t]
    \centering
    \includegraphics[trim={0pt 3pt 0pt 2pt}, clip, width=.9\linewidth]{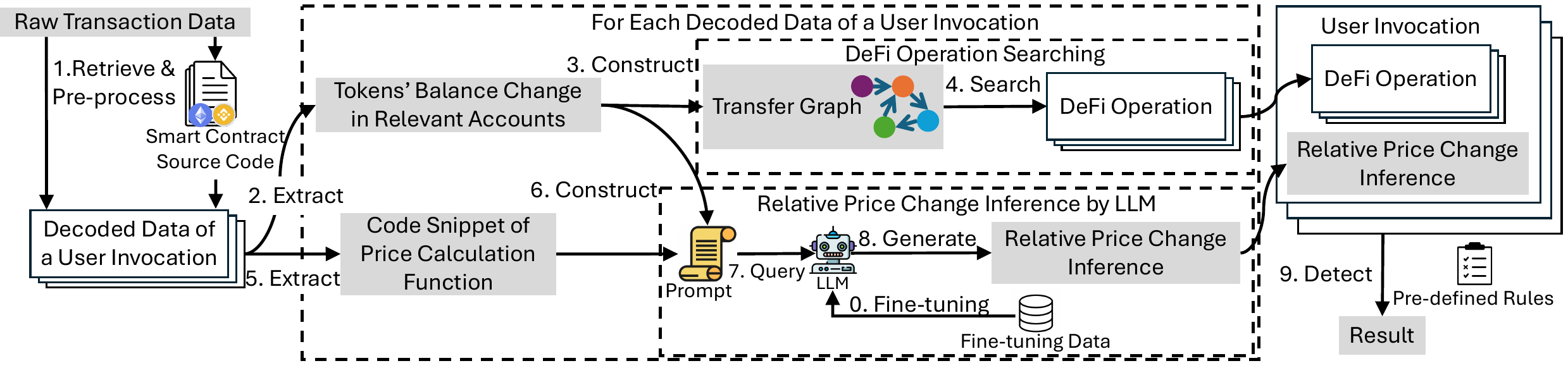}
    \vspace{-2ex}
    \caption{A high-level overview of \tool.
    }
    \label{fig:overview}
\end{figure*}

\begin{compactitem}
\item DeFort~\cite{DeFort24} uses historical exchange rates, i.e., prices, between two tokens to compute a so-called normal fluctuation range. 
However, it may be error-prone as historical prices, especially for low-liquidity or newly created pools, could vary significantly, thus being insensitive to the detection of subtle price manipulations. 
For instance, in the UwULend attack, slight manipulations observed in sUSDe prices --- decreasing by 4.2\,\%  or increasing by 4.43\,\%, did not exceed the predefined bounds that evades DeFort's detection,
illustrated in our evaluation in \mysec\ref{sec:evaluation}.

\item DeFiRanger~\cite{DeFiRanger23} detects abnormal price changes by tracing token exchange sequences within a transaction and comparing token exchange rates.
However, completely tracing these sequences is challenging for complex transactions. 
In the UwULend attack, the attacker crafted complicated deposit and withdrawal operations concerning WETH, making it difficult for DeFiRanger to detect.
\end{compactitem}

To address the aforementioned issues about price identification, 
our observation is that abnormal price fluctuations of tokens could be directly derived from the high-level price model and amount changes in token balances, without needing to explicitly calculate the exchange rate for a pair of tokens. 
Taking the UwULend attack as an example, swapping USDe into liquidity pools
decreases the value of $IP_{USDe,Pool_i}$, causing an abnormal drop in the median price, namely the price of sUSDe.
Based on this observation, in this paper, we propose a novel price change reasoning approach powered by LLM and integrate with predefined rules to enhance the capability of price manipulation attack detection.  


\section{Overview}
\label{sec:overview}

Based on the analysis of the motivating example illustrated in \mysec\ref{sec:motivatingexample}, we establish an intuition of inferring price change associated with DeFi operations during the transaction process to detect DeFi price manipulations in various scenarios.
While inferring through standard price models (i.e., CPMM and Stableswap as introduced in \mysec\ref{sec:backg_priceModel}) is straightforward, it is challenging to (i) interpret low-level price calculation Solidity code into high-level price calculation formulas, and (ii) infer the price change from the (custom) price calculation formulas and the information on the token balance changes in related accounts.
Both tasks require certain intelligence.

To address this key challenge, we introduce the first LLM-based approach, \name, for effective DeFi price manipulation detection.
As depicted in \myfig~\ref{fig:overview}, \name consists of ten steps.
In step \ding{172}, \name first decodes and slices raw transaction data, \fixme{and retrieves the source code of smart contracts called within the transaction}.
Then in steps \ding{173} \ding{176}
\name extracts the code of possible price calculation functions from smart contracts based on their signature, and token balance changes in relevant accounts.
Subsequently, in steps \ding{177}, \name embeds these two information into a prompt template, which will be used by the LLM to extract the price model and infer the price change in steps \ding{178} \ding{179}.
In the meantime, in steps \ding{174} \ding{175}, \name constructs the transfer graph (\mysec\ref{sec:transferGraph}) and uses it to recover the high-level DeFi operations (\mysec\ref{sec:searchDeFiOperation}) associated with those price change.
This is because using the trend of token price changes alone is insufficient for detecting price manipulation.
Finally, based on the recovered DeFi operations and their price change information, \name maps them into eight attack patterns listed in \mysec\ref{sec:finalDetect} and detects DeFi manipulation attacks in step \ding{180}.

We conduct an off-line fine-tuning step (step \textcircled{0})
to enhance LLMs' capabilities in extracting price calculation models and token price change reasoning.
This is because while the off-the-shelf LLMs may exhibit certain capability in reasoning and code understanding, they are limited for predicting the trend of token price changes given unlabeled code snippet and numerical changes of tokens balance, which will be illustrated in \mysec\ref{sec:RQ2}.
We now detail our fine-tuning technique in \mysec\ref{sec:priceInfer}. 

\section{Price Change Inference with LLMs}
\label{sec:priceInfer}
\definecolor{verylightgray}{rgb}{.97,.97,.97}

\lstdefinelanguage{Solidity}{
	keywords=[1]{anonymous, assembly, assert, balance, break, call, callcode, case, catch, class, constant, continue, constructor, contract, debugger, default, delegatecall, delete, do, else, emit, event, experimental, export, external, false, finally, for, function, gas, if, implements, import, in, indexed, instanceof, interface, internal, is, length, library, log0, log1, log2, log3, log4, memory, modifier, new, payable, pragma, private, protected, public, pure, push, require, return, returns, revert, selfdestruct, send, solidity, storage, struct, suicide, super, switch, then, this, throw, transfer, true, try, typeof, using, value, view, while, with, addmod, ecrecover, keccak256, mulmod, ripemd160, sha256, sha3}, 
	keywordstyle=[1]\color{blue}\bfseries,
	keywords=[2]{address, bool, byte, bytes, bytes1, bytes2, bytes3, bytes4, bytes5, bytes6, bytes7, bytes8, bytes9, bytes10, bytes11, bytes12, bytes13, bytes14, bytes15, bytes16, bytes17, bytes18, bytes19, bytes20, bytes21, bytes22, bytes23, bytes24, bytes25, bytes26, bytes27, bytes28, bytes29, bytes30, bytes31, bytes32, enum, int, int8, int16, int24, int32, int40, int48, int56, int64, int72, int80, int88, int96, int104, int112, int120, int128, int136, int144, int152, int160, int168, int176, int184, int192, int200, int208, int216, int224, int232, int240, int248, int256, mapping, string, uint, uint8, uint16, uint24, uint32, uint40, uint48, uint56, uint64, uint72, uint80, uint88, uint96, uint104, uint112, uint120, uint128, uint136, uint144, uint152, uint160, uint168, uint176, uint184, uint192, uint200, uint208, uint216, uint224, uint232, uint240, uint248, uint256, var, void, ether, finney, szabo, wei, days, hours, minutes, seconds, weeks, years},	
	keywordstyle=[2]\color{teal}\bfseries,
	keywords=[3]{block, blockhash, coinbase, difficulty, gaslimit, number, timestamp, msg, data, gas, sender, sig, value, now, tx, gasprice, origin},	
	keywordstyle=[3]\color{violet}\bfseries,
	identifierstyle=\color{black},
	sensitive=true,
	comment=[l]{//},
	morecomment=[s]{/*}{*/},
	commentstyle=\color{gray}\ttfamily,
	stringstyle=\color{red}\ttfamily,
	morestring=[b]',
	morestring=[b]"
}

\lstset{
	language=Solidity,
	backgroundcolor=\color{verylightgray},
	extendedchars=true,
	basicstyle=\footnotesize\ttfamily,
	showstringspaces=false,
	showspaces=false,
	numbers=left,
	numberstyle=\footnotesize,
	numbersep=9pt,
	tabsize=2,
	breaklines=true,
	showtabs=false,
	captionpos=b
}


\subsection{LLM Fine-tuning}
\label{sec:finetune}

For fine-tuning techniques, we \textit{by default} use OpenAI's fine-tuning paradigm~\cite{OpenAI_Fine-tuning} instead of supervised fine-tuning (SFT)~\cite{chung2024scaling, hu2021lora}, because the former requires only a small set of training data, whereas the latter generally requires more data points~\cite{ma2024combining}.
Nonetheless, we also demonstrate the \textit{generalizability} of \name's fine-tuning with open-source SFT models; see details in \mysec\ref{sec:discussion}.
In the default setting,
\name enhances the GPT-3.5-Turbo and GPT-4o models with data synthesized using the 
CPMM price model as illustrated in \mysec\ref{sec:backg_priceModel}, along with on-chain data to fine-tune them.

\fixme{Our fine-tuning process comprises two components: (a) on-chain data simulation, which generates realistic token balance-change pairs from transactions on a forked blockchain network; and (b) chain-of-thought (CoT)-style fine-tuning, where these simulated data are integrated into CoT-style prompts that are then used to fine-tune the LLM, enabling it to perform stepwise reasoning.} 

\noindent
\textbf{On-Chain Data Simulation.}
We leverage the fuzz testing method in Foundry~\cite{Foundry}, an off-the-shelf toolkit for Ethereum application development, to simulate on-chain data.
To avoid data leakage and generate a substantial volume of transactions satisfying the CPMM, we select the Uniswap V2:BTC20~\cite{UniswapBTC20} liquidity pool as our target.
We randomly generate inputs, namely integers ranging from $10^{20}$ to $10^{21}$ --- 100 Ether to 1000 Ether, for the swap operations which are simulated on a forked blockchain of block height 17,949,214.
Specifically, to include the data of inflating the price of tokens, we craft particular operations.
To begin with, we record the balance of WETH and BTC20 in the liquidity pool denoted as $bal_{WETH}$ and $bal_{BTC20}$ respectively.
Then we trigger \texttt{swapExactTokensForTokens} in contract \texttt{UniswapV2Router02}~\cite{UniswapRouter} to swap a amount of BTC20 for WETH, and record the latest balance of WETH, $bal_{WETH}^{'}$, and that of BTC20, $bal_{BTC20}^{'}$.
Finally, we obtained the tokens' balance change as a pair ($bal_{WETH} - bal_{WETH}^{'}, bal_{BTC20} - bal_{BTC20}^{'}$).
In terms of deflating the price of tokens, we swap a amount of WETH for BTC20 instead, with similar subsequent operations.
Finally, we build a fine-tuning database 
with 500 pairs each for price inflating and deflating.
Despite the only use of CPMM-based DeFi protocols, our evaluation results in~\mysec\ref{sec:evaluation} demonstrates a significant gain in term of price manipulation attack detection for DeFi protocols using custom price models.


\begin{figure}[!t]
    \centering
    \includegraphics[trim={0 5pt 0 6pt}, clip, width=1.0\linewidth]{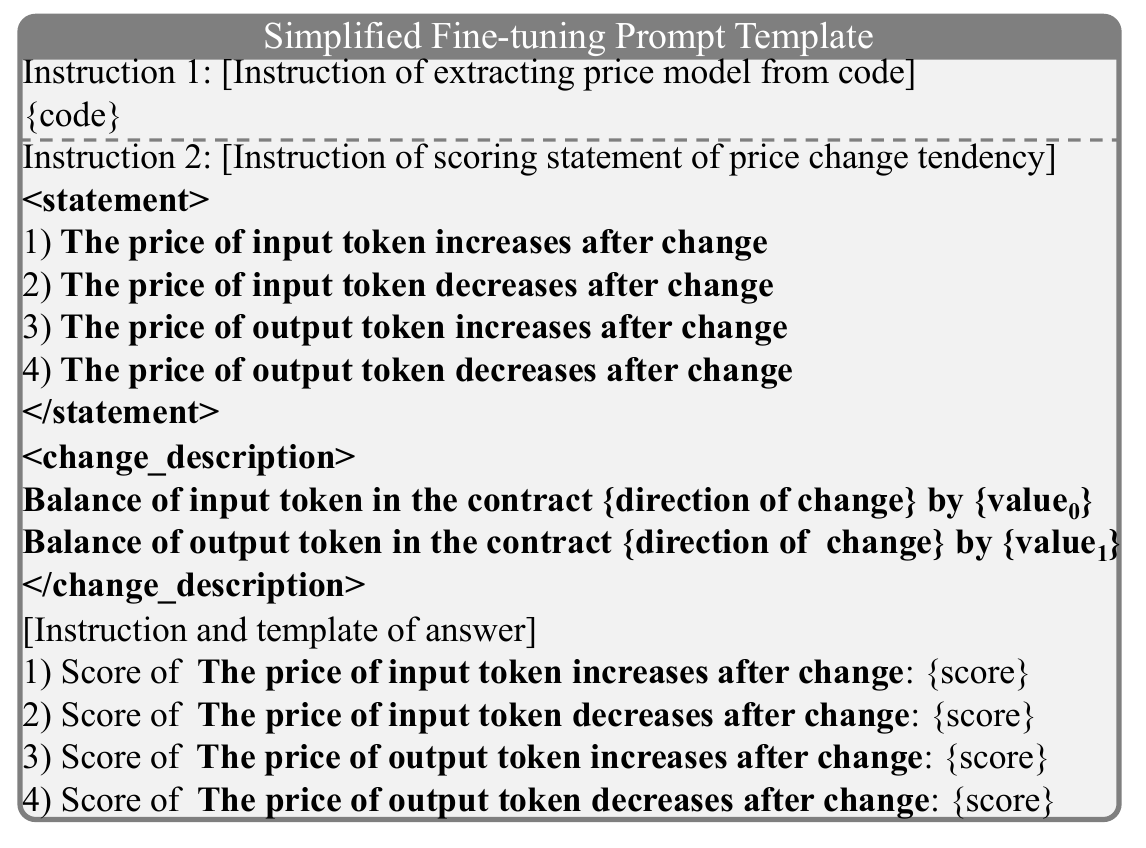}
    \vspace{-5ex}
    \caption{The simplified prompt template used in fine-tuning the LLM. The full version can be found in our supplementary material~\cite{supplementary_material}.}
    \label{fig:fine-tuning_prompt}
\end{figure}

\noindent
\textbf{CoT-style Fine-tuning.}
\myfig\ref{fig:fine-tuning_prompt} demonstrates the prompt template used in the fine-tuning.
\fixme{We carefully construct a CoT-style fine-tuning prompt for guiding the LLM toward step-by-step reasoning, which}
integrates both
\fixme{simulated on-chain data collected in the previous step}
and the price context.
Above the dashed line is the first instruction, which requires the LLM to extract the price calculation model from the provided code.
\texttt{\{code\}} is the placeholder for the code snippet of price calculation functions.
Below the dashed line, we guide the LLM to evaluate the credibility of four statements based on the price model extracted from step 1 and the tokens' balance change.
We demand that the LLM expresses the credibility of a statement using integers ranging from 1 to 10.
Compared to merely responding with a simple ``Yes'' or ``No'', this scoring method also indicates the confidence level of the responses, which can assist us in selecting the answers in which the LLM is more confident. 
By offering a finer-grained evaluation scale, it reveals when the model is uncertain—if two contradictory statements receive the same confidence score, we know it cannot distinguish which is more trustworthy and therefore discard those low-confidence results.

In the data part of the template, \texttt{\{$value_0$\}} and \texttt{\{$value_1$\}} are the placeholders for the first and second values in a price change pair, which are sampled from the fine-tuning dataset, respectively.
\texttt{\{direction of change\}} can be either ``increases'' or ``decreases.''
Specifically, if \texttt{\{$value_i$\}} is greater than 0, \texttt{\{direction of change\}} is ``increases''; conversely, if \texttt{\{$value_i$\}} is less than 0, \texttt{\{direction of change\}} is ``decreases.''
For the answer part, \texttt{\{score\}} is the score placeholder, an integer between 1 and 10.


Following OpenAI's fine-tuning guideline~\cite{OpenAI_Fine-tuning} that recommends using 50 to 100 training examples, we randomly sample 96 non-repetitive (in current and previous training sets) data from the fine-tuning database, and allocate 83\% of the samples for training and 17\% for validation.
Subsequently, we insert data from the training and validation sets into the prompt template.
To obtain the desired response for each prompt, we firstly ask LLMs to generate raw response, including the analysis of price model and scores of statements, for the given prompt. 
Next, we manually verify the correctness of the responses, of which the correct responses are stored and the wrong responses will also be corrected. Particularly, we simply swap the scores of two opposite statements to correct the error in them.
Through this process, we construct a ground truth about prompts and its responses for fine-tuning. 

During the fine-tuning process, training hyperparameters were automatically configured by the OpenAI fine-tuning API (for SFT fine-tuning, hyperparameters were manually set using common practice; see \mysec\ref{sec:discussion}).
In particular, once the model achieved 100\% accuracy on the validation set, we terminated the training to avoid overfitting. The entire fine-tuning process is cost-efficient, consuming a total of around 1 million training tokens for each model.
The detailed costs across different models will be introduced in \mysec\ref{sec:RQ2}.



\begin{figure*}
   \centering
   \includegraphics[trim={5pt 12pt 5pt 8pt}, clip, width=1\linewidth]{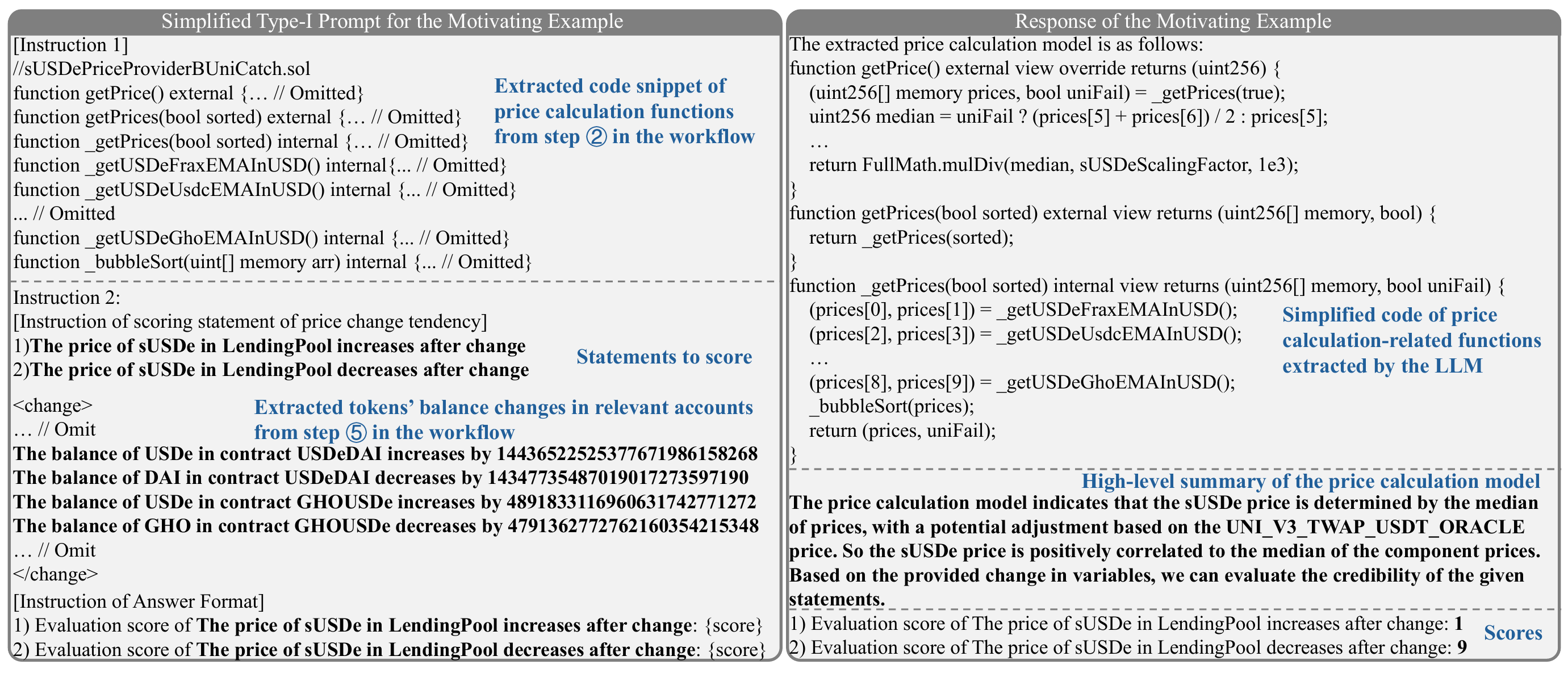}
   \caption{\fixme{The simplified Type-I prompt and its response for the motivating example during the inference process.}
   \label{fig:moti_prom_res}}
\end{figure*}

\subsection{Inference with the Fine-tuned LLM}
\label{sec:inference}

We first illustrate the \emph{general inference process} using the motivating example.
Nevertheless, there could be closed source DeFi protocols that dissatisfy steps \ding{176}\ding{177} so that we design a \emph{customized inference process} to deal with them.

\noindent
\textbf{General Inference Process.}
To assist LLMs in extracting and analyzing the price model, our inference prompt provides relevant code for LLMs to score the trend of price changes and guides them in inferring the token price changes.
\fixme{\myfig~\ref{fig:moti_prom_res} illustrates a simplified version of the prompt used and the response produced by our fine-tuned LLM for inferring price changes of the motivating example in \mysec\ref{sec:motivatingexample}.}
\fixme{We call this Type-I prompt,}
\fixme{used for the typical cases with code input retrieved from steps \ding{176}\ding{177}.}
It is different from the fine-tuning prompt in the 
\texttt{<statement>}
and 
\texttt{<change\_description>}
parts.
In the fine-tuning prompt shown in \myfig~\ref{fig:fine-tuning_prompt}, both parts are fixed, while they are dynamically generated during inference using two formats:
(i) ``The price of \texttt{\{token\_name\}} in \texttt{\{contract\_name\}} \texttt{\{direction\_of\_change\}} after change'' for the 
part \texttt{<statement>},
and (ii) ``The balance of \texttt{\{token\_name\}} in \texttt{\{contract\_name\}} \texttt{\{direction\_of\_change\}} by \texttt{\{change\_value\}}'' or ``The total supply of \texttt{\{token\_name\} \{direction\_of\_change\}} by \texttt{\{change\_value\}}'' for the 
part \texttt{<change\_description>}.
Specifically,
to fill them in Type-I prompt, \name generates a pair of statements for each token, i.e., one regarding the increase in token price and another regarding the decrease in token price.

\name asks the fine-tuned LLM to locate the price calculation model from the input code and evaluate the credibility of the generated statements.
\fixme{From the motivating example's response shown in the right-hand section of \myfig~\ref{fig:moti_prom_res},}
\fixme{the LLM first extracts the code of price calculation-related functions, followed by a high-level summary.}
\fixme{In this example, it accurately identifies the underlying price model}
\fixme{(see \Cref{PriceOfsUSDe}):}
\fixme{the price of \texttt{sUSDe} is determined by the median of multiple prices.}
\fixme{With this knowledge, the LLM can evaluate two opposing statements based on the changes in token balances and provide the correct answer with}
\fixme{high confidence, i.e., a higher score.}


\noindent
\textbf{Customized Inference Process.}
Although the majority of DeFi applications are open source to gain users' trust, some remain closed source, making our Type-I prompt inapplicable.
To address this, we developed a Type-II prompt template
to infer the trend of price changes in closed source two-token liquidity pools.
Our observation is that the majority of two-token liquidity pools use the CPMM as their underlying price model.
Therefore, the primary distinction between the Type-I prompt and the Type-II prompt lies in replacing the Instruction 1 with a description of the liquidity pool, informing the LLM that the pool's price model aligns with CPMM.
It is worth noting that the liquidity pool is automatically identified during transaction analysis, which will be introduced in \mysec\ref{sec:defiOperation}.
\fixme{To evaluate its effectiveness, we applied Type-II on a test dataset~\cite{ds_ft_test_set} of 100 price changes on CPMM, and it reached 97\% accuracy, close to the 99\% accuracy of Type-I. This demonstrates that, if applied to previously identified two-token liquidity pools, which commonly adopt CPMM, Type-II is substantially effective comparable to Type-I, while also adapting to closed-source pools.}
More details about the Type-II prompt and a case study of using it are available in~\cite{supplementary_material}.



\section{DeFi Operations}
\label{sec:defiOperation} 

The standalone fluctuations in token prices are meaningless; they need to be considered within the DeFi context to serve as evidence for detecting price manipulation.
However, the raw transactions obtained from the blockchain consist solely of low-level information, such as token transfer actions and smart contract invocations.
There exists a gap between raw transactions and high-level DeFi semantics.

To bridge this gap, we first model token transfer actions using a directed graph (\mysec\ref{sec:transferGraph}), and then recover high-level DeFi operations from it (\mysec\ref{sec:searchDeFiOperation}).
Since our detection is based on a single transaction, it should be noted that all described operations are derived from one raw transaction, and we do not consider the DeFi operations expressed by the combination of multiple raw transactions.
Based on our study of the top-10 high-value DeFi applications 
(the full list could be found in~\cite{supplementary_material}) across
three categories --- Decentralized EXchange (DEX), Lending, Yield Farming,
with active transactions in each category, due to the susceptibility to front-running across multiple transactions and the atomicity of transactions ensuring complete execution of operations, only a very few DeFi operations span multiple transactions.

\subsection{Transfer Graph Construction}
\label{sec:transferGraph}

We define the Transfer Graph (TG) (\mydef~\ref{def:TG}), a directed graph where the edges represent transfer actions (\mydef~\ref{def:transfer}) and the vertices represent related accounts, to model transfer actions within each user invocation.
\begin{definition}[Transfer]
\label{def:transfer}
    A transfer $T := \langle s, r, t, v \rangle$, if performed successfully, deducts amount $v \in \mathbb{N}$ of token $t \in Addr$ from the sender's account $s \in Addr$ and the balance of token $t$ in the receiver's account $r \in Addr$ increases by $v$.
\end{definition}
\begin{definition}[Transfer Graph]
\label{def:TG}
    A \textit{Transfer Graph} (\textit{TG}) is a tuple ($\mathcal{A}$,$\mathcal{E}$), 
    where $\mathcal{A}$ is the set of all accounts (including EOAs, CAs and $\emptyset$) involved in a user invocation, 
    $\mathcal{E}$ is the set of directed edges, i.e., $\mathcal{E} = \{E_1,...,E_m\}\subseteq\mathcal{A}\times\mathcal{A}$, where each $E_i := \langle j, T_k \rangle$, j is the time index of $T_k$, $T_k \in \mathcal{T}$, 
    $\mathcal{T}$ is the set of all transfer actions involved in the user invocation, i.e., $\mathcal{T} = \{T_1,...,T_n\}$, where $T_i.s, T_i.r \in \mathcal{A}$ for each $T_i$.
\end{definition}
    

According to our categorization, a transfer action can be one of three types: 
\textit{transferring}, \textit{burning}, and \textit{minting token}.
All transfer actions can be expressed as ``\textit{Sender} transfers \textit{amount} of \textit{token} to \textit{Receiver}.''
In the \textit{transferring token},
all accounts involved must be either EOAs or CAs, and must not be a zero address
or a dead address
(we uniformly denote these two special addresses by $\emptyset$).
Meanwhile, the \textit{receiver} in \textit{burning token} and the \textit{sender} in \textit{minting token} actions must be $\emptyset$.

\begin{figure}[!t]
    \centering
    \includegraphics[trim={0 15pt 0 5pt}, clip, width=1\linewidth]{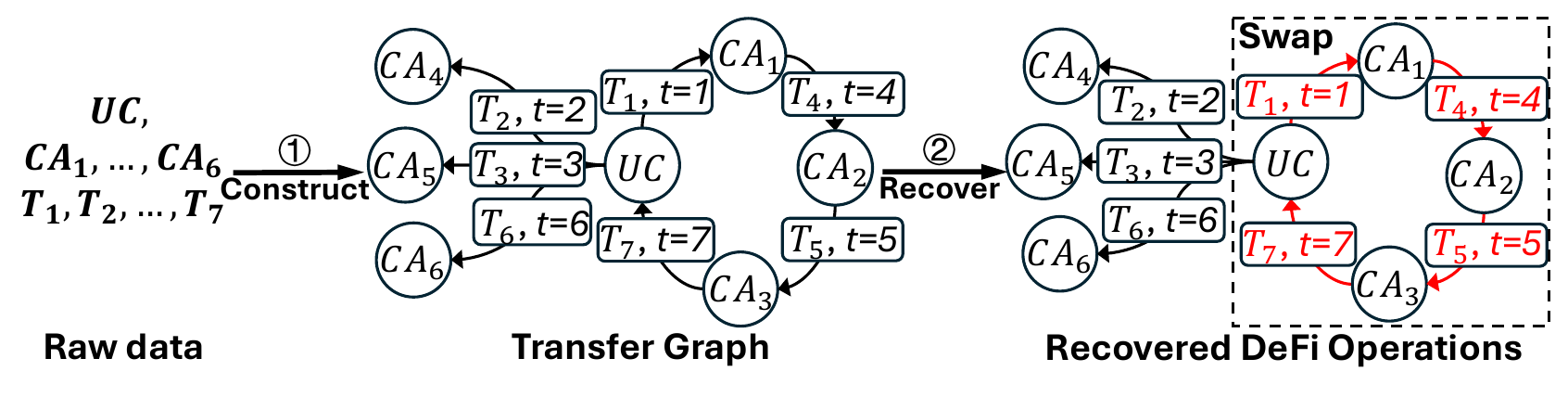}
    \vspace{-4ex}
    \caption{An illustrative example showing the workflow of recovering DeFi operations. $UC$: a collection of user-controlled accounts, including EOAs and CAs; $CA_i$: the $i$-th contract account; $T_i$: the $i$-th \textit{transferring token} action.}
    \label{fig:TG}
\end{figure}

\myfig~\ref{fig:TG}, as demonstrated in step \ding{172}, illustrates the construction of a TG from the raw transaction of a user invocation. This user invocation includes six contract accounts and a collection of user-controlled accounts $UC$, which includes EOAs and CAs, along with seven \textit{transferring token} actions.
The \textit{Sender} and \textit{Receiver} of a transfer action are connected by a directed edge, from the \textit{Sender} to the \textit{Receiver}, with a time index to indicate the order of occurrence.
$T_1$, from one of the user-controlled accounts in $UC$ to $CA_1$, is the first transfer action in this user invocation.
Before $CA_1$ transfers tokens to $CA_2$ through $T_4$, user-controlled accounts initiate two transfers to $CA_4$ and $CA_5$ respectively, resulting in $T_4$ having a larger time index compared to $T_2$ and $T_3$.
Similarly, since $T_6$ occurs between $T_5$ (with a time index of 5) and $T_7$ (with a time index of 7), its time index is set to 6.

\fixme{Compared to the CFT approach used in DeFiRanger~\cite{DeFiRanger23}, which models invocation and transfer actions within raw transactions, our TG provides finer granularity. TG represents transfers as a directed graph while preserving their temporal order, enabling it to capture more complex operations more effectively. In our evaluation on 1,000 real-world Ethereum transactions, TG achieved a TPR ($\frac{\#TP}{\#TP+\#FN}$) of 0.912, significantly higher than CFT's 0.559, while maintaining comparable precision ($\frac{\#TP}{\#TP+\#FP}$), i.e., 0.984 vs. 0.991. TG further demonstrated substantial advantages in recovering complex swap operations involving multiple relayers, achieving an 89.3\% improvement over CFT. More details about the experimental setup and analysis are available in ~\cite{supplementary_material}.
Compared to prior fund-flow graph approaches, such as ~\cite{mclaughlin2023large}, TG differs in two key aspects: (i) it is constructed at the level of user invocations rather than transactions, which may contain multiple invocations, and (ii) it is designed for high-level DeFi operation recovery, whereas those primarily target behavior analysis (e.g., arbitrage).}

\subsection{DeFi Operation Inference}
\label{sec:searchDeFiOperation}

\begin{table*}[!t]
    \centering
    \caption{Price change information-directed attack patterns.
    }
    \vspace{-2ex}
    \begin{tabular}{ccl}
         \toprule
         Type&\multicolumn{1}{c}{Pattern}&\multicolumn{1}{c}{Details}\\
         \midrule
         \multirow{2}{5em}[-0.8em]{\makecell[c]{Buy\\\&\\Sell\\~\cite{elephantMoney}}}&I& \makecell[l]{1) Swap $Token_x$ to $Token_y$ through $Pool_{buy}$\\2) The price of $Token_y$ in  $Pool_{sell}$ increases / The price of $Token_z$ in $Pool_{sell}$ decreases\\3) Swap $Token_y$ to $Token_z$ through $Pool_{sell}$}\\
         \cline{2-3}
         &II& \makecell[l]{1) The price of $Token_x$ in $Pool_{buy}$ increases / The price of $Token_y$ in $Pool_{buy}$ decreases/\\ \ \ \ \ The price of $Token_y$ in $Pool_{sell}$ increases / The price of $Token_z$ in $Pool_{sell}$ decreases\\2) Swap $Token_x$ to $Token_y$ through $Pool_{buy}$\\3) Swap $Token_y$ to $Token_z$ through $Pool_{sell}$}\\
         \hline
         \multirow{2}{5em}[-0.8em]{\makecell[c]{Deposit\\\&\\Borrow\\~\cite{CreamFinance}}}&III& \makecell[l]{1) Deposit $Token_x$ into $Contract_{deposit}$ and get $Token_y$ as credential\\2) The price of $Token_x$ in $Contract_{borrow}$ increases / The price of $Token_z$ in $Contract_{borrow}$ decreases\\3) Borrow $Token_z$ using $Token_x$ as collateral from $Contract_{borrow}$}\\
         \cline{2-3}
         &IV& \makecell[l]{1) The price of $Token_x$ in $Contract_{borrow}$ increases / The price of $Token_z$ in $Contract_{borrow}$ decreases\\2) Deposit $Token_x$ into $Contract_{deposit}$ and get $Token_y$ as credential\\3) Borrow $Token_z$ using $Token_x$ as collateral from $Contract_{borrow}$}\\
         \hline
         \multirow{2}{5em}[-0.8em]{\makecell[c]{Stake\\\&\\Claim\\~\cite{ATK}}}&V& \makecell[l]{1) Stake $Token_x$ into $Contract_{stake}$\\2) The price of $Token_y$ in $Contract_{claim}$ decreases\\3) Claim $Token_y$ from $Contract_{claim}$}\\
         \cline{2-3}
         &VI& \makecell[l]{1) The price of $Token_x$ in $Contract_{stake}$ increases / The price of $Token_y$ in $Contract_{claim}$ decreases \\2) Stake $Token_x$ into $Contract_{stake}$\\3) Claim $Token_y$ from $Contract_{claim}$}\\
         \hline
         \multirow{2}{5em}[-0.8em]{\makecell[c]{Deposit\\\&\\Withdraw\\~\cite{Harvest}}}&VII& \makecell[l]{1) Deposit $Token_x$ into $Contract_{deposit}$ and get $Token_y$ as credential\\2) The price of $Token_y$ in $Contract_{withdrawal}$ increases / The price of $Token_z$ in $Contract_{withdrawal}$ decreases \\3) Withdraw $Token_z$ from $Contract_{withdrawal}$ by burning $Token_y$}\\
         \cline{2-3}
         &VIII& \makecell[l]{1) The price of $Token_x$ in $Contract_{deposit}$ increases / The price of $Token_y$ in $Contract_{deposit}$ decreases /\\ \ \ \ \ The price of $Token_y$ in $Contract_{withdrawal}$ increases / The price of $Token_z$ in $Contract_{withdrawal}$ decreases\\2) Deposit $Token_x$ and get $Token_y$ as credential\\3) Withdraw $Token_z$ from $Contract_{withdrawal}$ by burning $Token_y$}\\
         \bottomrule
    \end{tabular}
    \label{tab:attack_pattern}
\end{table*}

Based on our in-depth analysis of 
the top-30 DeFi apps,
we identify six DeFi operations that need to be recovered:
1) \textit{Swap}, primarily from DEXs such as Uniswap; 
2) \textit{Deposit}, defined by yield-farming and lending apps such as AAVE and Pendle; 
3) \textit{Withdraw}, mainly from yield-farming apps; 
4) \textit{Borrow}, based on lending protocols like Compound; 
5) \textit{Stake} and 6) \textit{Claim}, both predominantly from yield-farming apps like Convex Finance.

\textbf{Swap} involves a user exchanging one token ($Token_{in}$) for another ($Token_{out}$) from a liquidity pool. Relevant contract accounts, excluding user-controlled ones, are noted as liquidity pools for price trend analysis when source code is unavailable.
In a \textbf{Deposit}, a user transfers a token ($Token_{deposit}$) to a yield-farming or lending protocol and receives a proof token ($Token_{proof}$) through minting. 
\textbf{Withdraw} occurs when a user retrieves tokens ($Token_{withdraw}$) from a protocol by burning a proof token ($Token_{proof}$).
\textbf{Borrow} refers to a DeFi operation where a borrower receives a token ($Token_{borrow}$) and incurs a debt evidenced by another token ($Token_{debt}$), issued through minting.
\textbf{Stake} happens in yield-farming protocols like Convex Finance~\cite{convex_fi}, where users deposit tokens and can earn rewards without receiving a minted proof token.
\textbf{Claim} enables users to retrieve staked tokens ($Token_{claim}$) and bonuses without burning any tokens, unlike the \textit{Withdraw} operation which requires burning tokens to retrieve assets.

\noindent\textbf{An Operation Recovering Example.}
To recover DeFi operations from the TG, we design and employ a search algorithm based on directed graphs for each operation.
%
\myfig~\ref{fig:TG} illustrates that, to recover a \textit{Swap} operation, we use the depth-first search algorithm to identify cycles that both start and end at user-controlled accounts.
These cycles must satisfy three constraints:
(i) the transfer action of each edge must be \textit{transferring token};
(ii) the time index of each edge must be monotonically increasing;
and (iii) the token involved in the first transfer action must differ from the token in the last transfer action.
Therefore, from the well-constructed TG, we can recover the \textit{Swap} operation expressed as $UC \rightarrow T_1 \rightarrow CA_1 \rightarrow T_4 \rightarrow CA_2 \rightarrow T_5 \rightarrow CA_3 \rightarrow T_7 \rightarrow UC$, and we label $CA_1$, $CA_2$, and $CA_3$ as liquidity pools.

\section{Price Manipulation Detection}
\label{sec:finalDetect}


The price change information and high-level DeFi operations recovered from \mysec\ref{sec:priceInfer} and \mysec\ref{sec:defiOperation}, respectively, are finally checked against the detection rules (c.f. Table~\ref{tab:attack_pattern}).
We analyze 
four prevalent
types of DeFi protocols 
and their associated attack instances.
Specifically, we examine ElephantMoney~\cite{elephantMoney} for \textit{DEX}, Cream Finance~\cite{CreamFinance} for \textit{lending protocols}, ATK~\cite{ATK} for \textit{staking-based yield farming protocols}, and Harvest~\cite{Harvest} for \textit{deposit-based yield farming protocols}, which collectively incurred a loss of \$163.3M.

Based on this in-depth analysis, we identify four attack types targeting different DeFi protocols and their eight generalized attack patterns, as depicted in Table~\ref{tab:attack_pattern}.
These attack types are \textit{Buy \& Sell}, \textit{Deposit \& Borrow}, \textit{Stake \& Claim}, and \textit{Deposit \& Withdraw}, with each type corresponding to two specific attack patterns.
More details about these attack types and patterns are available in our supplementary material~\cite{supplementary_material}.

We implement \tool with about 3,900 lines of Python code.
\tool currently supports two blockchains, i.e., Ethereum~\cite{ethereum} and BSC~\cite{bsc}, which account for over 60\% of total value locked (TVL) among all the blockchains~\cite{defillama_chains}.
To obtain raw data from specific transactions, we utilize blockchain node APIs facilitated by QuickNode~\cite{quicknode}, an external RPC service provider.

\section{Evaluation}
\label{sec:evaluation}


We aims to evaluate the following research questions (RQs):
\begin{itemize}[leftmargin=*,noitemsep,topsep=0pt]
    \item \textbf{RQ1}: How effectively does \tool detect price manipulation attacks compared with the existing state-of-the-arts?
    \item \textbf{RQ2}: How significantly does the fine-tuning technique promote the accuracy of \tool?
    \item \textbf{RQ3}: How practically and efficiently does \tool detect price manipulation attacks in a real-world setting? 
\end{itemize}



\noindent
\textbf{Datasets.}
To address these RQs, we collect three datasets from real DeFi transactions on Ethereum and BSC to evaluate \name, as shown in~\Cref{tab:benchmark}.
Specifically, we use the first dataset \textit{D1}, which comprises 95 transactions of real-world price manipulation attacks from 90 DeFi applications, to evaluate both RQ1 and RQ2.
To demonstrate \name's practicality for RQ3, we use the second dataset \textit{D2}, consisting of 968 suspicious transactions collected by our industry partner.
Furthermore, to measure \name's time overhead when deployed in a realistic setting with both suspicious and benign transactions, we mix 968 suspicious transactions from \textit{D2} with 96,800 benign transactions in the third dataset \textit{D3}.

For \textit{D1}, we scraped data from multiple sources.
Initially, we included all 54 price manipulation transactions identified by DeFort~\cite{DeFort24}.
We also collected all 55 price manipulation attacks documented by the renowned DeFi security GitHub repository, DeFiHackLabs~\cite{DeFiHackLabs}, from October 26, 2020, to October 11, 2023.
To further expand our dataset, we acquired 31 transactions confirmed as price manipulation attacks from our industry partner.
After removing duplicates among the three sources of data, we ultimately obtained 95 price manipulation attacks for \textit{D1}, which collectively caused \$381.16M in losses on Ethereum and BSC.
We also measure the distribution and monetary losses of the 95 attacks in \textit{D1} across our eight attack patterns described in \mysec\ref{sec:finalDetect}, as shown in Table~\ref{tab:distr_and_loss}.

\textit{D2} comprises 968 suspicious transactions provided by our industry partner, a Web3 security company, which monitors blockchain transactions in real-time and automatically flags transactions that yield significant profits for initiators.
All these transactions are sufficiently complex and potentially involve price manipulation.
While manually confirming these transactions is labor-intensive and prone to errors, 155 of them have been confirmed to belong to other vulnerabilities, such as Reentrancy (OpenLeverage~\cite{repo_openLeverage}), Unverified User Input (YIELD~\cite{repo_yield}), and Access Control Bugs (SafeMoon~\cite{repo_safeMoon}), by developers and industry partners.
In this way, we can stress-test \name's detection capabilities when mixed with other types of attack transactions. 

Furthermore, to assess \name's false alarms on benign transactions and to measure its time overhead in a realistic setting, we need to collect a large number of benign transactions, as the most majority of real-world transactions are still benign.
In the absence of empirical data on the ratio of suspicious to benign transactions, we adopt a conservative yet reasonable ratio of 1:100, assuming one suspicious transaction for every 100 benign transactions.
Thus, with 968 suspicious transactions in \textit{D2}, we require 96,800 benign transactions for \textit{D3}.
To this end, we randomly sampled these 96,800 transactions from DeFort~\cite{DeFort24}'s dataset
of 428,523 benign transactions, which includes 384,143 benign transactions on Ethereum and BSC.
At this sample size, we achieve a 99.999\% confidence level with a margin of error of 0.625\%.



\begin{table}[t!]
    \centering
    \small
    \caption{The benchmark datasets used for evaluation.}
    \vspace{-2ex}
    \begin{tabular}{ll}
    \toprule
        Dataset & RQs \\
        \midrule
        D1: 95 historical real-world attacks & RQ1, RQ2\\
        \midrule
        D2: 968 suspicious transactions & RQ3\\
        \midrule
        D3: 96,800 benign transactions & RQ3\\
    \bottomrule
    \end{tabular}
    \label{tab:benchmark}
\end{table}

\begin{table}[!t]
    \centering
    \small
    \caption{Summary of different attack patterns in \textit{D1}.}
    \vspace{-2ex}
    \resizebox{\columnwidth}{!}{
        \begin{tabular}{lcccccccc}
        \toprule
            Pattern & I & II & III & IV & V & VI & VII & VIII
            \\
            \midrule
            \#Case & 49 & 20 & 5 & 6 & 1 & 2 & 11 & 1\\
            Loss(\$) & 55.3M & 70M & 141.2M & 43M & 61K & 83K & 71.4M & 9K\\
            \bottomrule
        \end{tabular}
    }
    \label{tab:distr_and_loss}
\end{table}

\noindent
\textbf{Minimized Data Leakage.}
\tool
utilizes LLMs to assess token price changes solely by 
using
price calculation functions and token balance changes, without any knowledge of attack transactions.
Even if related attacks appeared in pre-training,
evaluating price change tendencies in this manner falls outside its training data, resulting in no or minimized data leakage.

\noindent
\textbf{Experimental Setup.}
All experiments were conducted on a desktop computer running Ubuntu 20.04, powered by an Intel® Xeon® W-2235 CPU (3.80 GHz, 6 cores, and 12 threads) and equipped with 16 GB of memory.
For LLMs, we \textit{by default} use OpenAI's GPT family models
but also explore open-source SFT model tuning in \mysec\ref{sec:discussion},
as explained in \mysec\ref{sec:finetune}.
\name by default uses GPT-3.5-Turbo (\texttt{GPT-3.5-turbo-1106}) for its well-recognized cost-performance balance, but we also conduct an ablation study in \mysec\ref{sec:RQ2} to test the more advanced GPT-4o (\texttt{GPT-4o-2024-08-06}) for fine-tuning.
For the LLM configuration, \name employs nucleus sampling~\cite{holtzman2019curious} with a \texttt{top-p} value of 1 and sets the \texttt{temperature} to 0, ensuring highly deterministic responses for each prompt, although each was run only once.

\begin{table*}[!t]
    \centering
\caption{Detection results for 95 ground-truth DeFi price manipulation attacks. 
} 
    \vspace{-2ex}
    \scriptsize
    \setlength{\tabcolsep}{4pt}
    \begin{minipage}{0.48\linewidth}
    \resizebox{\textwidth}{!}{
       \begin{tabular}{llllllll}
    \toprule
        Protocol & Chain & Date & Loss & DS & DT & DF & DR \\ \midrule
        AES & BSC & 07-Dec-22 & 60K & \cmark & \xmark & \cmark & \xmark  \\
        APC & BSC & 01-Dec-22 & 6K & \cmark & \xmark & \cmark & \xmark \\ 
        APEDAO & BSC & 18-Jul-23 & 7K & \cmark & \cmark & \xmark & \cmark \\ 
        ApeRocket & BSC & 14-Jul-21 & 1.26M & \textcircled{1} & \cmark & \cmark & \xmark \\ 
        ARK & BSC & 23-Mar-24 & 201K & \xmark & \xmark & \xmark & \xmark \\ 
        ArrayFinance & ETH & 18-Jul-21 & 516K & \textcircled{2} & \xmark & \xmark & \cmark \\ 
        ATK & BSC & 12-Oct-22 & 61K & \xmark & \xmark & \cmark & \xmark \\ 
        AutoSharkFinance\_1 & BSC & 29-Oct-21 & 2M & \cmark & \cmark & \cmark & \cmark \\ 
        AutoSharkFinance\_2 & BSC & 24-May-21 & 750K & \cmark & \cmark & \cmark & \xmark \\ 
        BabyDoge & BSC & 28-May-23 & 137K & \cmark & \xmark & \xmark & \xmark \\ 
        Bamboo & BSC & 04-Jul-23 & 117K & \cmark & \xmark & \xmark & \cmark \\ 
        BBOX & BSC & 16-Nov-22 & 12K & \cmark & \xmark & \cmark & \xmark \\ 
        BDEX & BSC & 30-Oct-22 & 3K & \textcircled{2} & \cmark & \cmark & \xmark \\ 
        bDollar & BSC & 21-May-22 & 730K & \cmark & \xmark & \xmark & \cmark \\ 
        BEARNDAO & BSC & 05-Dec-23 & 769K & \cmark & \xmark & \cmark & \cmark \\ 
        BeltFinance\_1 & BSC & 29-May-21 & 408K & \cmark & \cmark & \xmark & \cmark \\ 
        BeltFinance\_2 & BSC & 29-May-21 & 6.23M & \cmark & \cmark & \xmark & \cmark \\ 
        BFCToken & BSC & 09-Sep-23 & 38K & \cmark & \xmark & \xmark & \cmark \\ 
        BGLD & BSC & 12-Dec-22 & 18K & \cmark & \cmark & \cmark & \xmark \\ 
        BH & BSC & 11-Oct-23 & 1.27M & \cmark & \cmark & \xmark & \cmark \\ 
        BTC20 & ETH & 19-Aug-23 & 47K & \cmark & \xmark & \xmark & \xmark \\ 
        BXH & BSC & 28-Sep-22 & 40K & \cmark & \cmark & \cmark & \xmark \\ 
        bZx & ETH & 18-Feb-20 & 350K & \cmark & \xmark & \cmark & \cmark \\ 
        Carson & BSC & 26-Jul-23 & 150K & \textcircled{1} & \xmark & \cmark & \cmark \\ 
        Cellframe & BSC & 01-Jun-23 & 76K & \cmark & \cmark & \xmark & \cmark \\ 
        CheeseBank & ETH & 06-Nov-20 & 3.3M & \cmark & \cmark & \cmark & \cmark \\ 
        ConicFinance & ETH & 21-Jul-23 & 3.25M & \cmark & \xmark & \xmark & \cmark \\ 
        CreamFinance & ETH & 27-Oct-21 & 130M & \cmark & \cmark & \cmark & \xmark \\ 
        CS & BSC & 23-May-23 & 714K & \cmark & \cmark & \xmark & \cmark \\ 
        Cupid & BSC & 31-Aug-22 & 78K & \cmark & \cmark & \cmark & \cmark \\ 
        DFS & BSC & 30-Dec-22 & 2K & \cmark & \xmark & \cmark & \cmark \\ 
        Discover & BSC & 06-Jun-22 & 11K & \cmark & \xmark & \xmark & \xmark \\ 
        DotFinance & BSC & 25-Aug-21 & 430K & \cmark & \cmark & \cmark & \xmark \\ 
        EAC & BSC & 29-Aug-23 & 17K & \cmark & \xmark & \cmark & \cmark \\ 
        EGDFinance & BSC & 07-Aug-22 & 36K & \cmark & \cmark & \cmark & \xmark \\ 
        ElephantMoney & BSC & 12-Apr-22 & 11.2M & \cmark & \xmark & \cmark & \xmark \\ 
        Eminence & ETH & 29-Sep-20 & 7M & \cmark & \cmark & \cmark & \xmark \\ 
        ERC20TokenBank & ETH & 31-May-23 & 111K & \cmark & \xmark & \xmark & \cmark \\ 
        FFIST & BSC & 19-Jul-23 & 91K & \cmark & \xmark & \cmark & \xmark \\ 
        GDS & BSC & 03-Jan-23 & 180K & \cmark & \cmark & \cmark & \xmark \\ 
         GPT & BSC & 24-May-23 & 42K & \cmark & \xmark & \cmark & \xmark \\ 
        Groker20 & ETH & 10-Nov-23 & 68K & \cmark & \xmark & \xmark & \cmark \\ 
        GymDeFi & BSC & 09-Apr-22 & 312K & \cmark & \xmark & \cmark & \cmark \\ 
        Hackerdao & BSC & 24-May-22 & 65K & \cmark & \xmark & \cmark & \cmark \\ 
         Harvest & ETH & 26-Oct-20 & 21.5M & \textcircled{2} & \cmark & \cmark & \cmark \\ 
        IndexedFinance & ETH & 14-Oct-21 & 16M & \xmark & \xmark & \xmark & \cmark \\ 
        INUKO & BSC & 14-Oct-22 & 50K & \xmark & \xmark & \cmark & \xmark \\ 
        InverseFinance\_1 & ETH & 16-Jun-22 & 1.26M & \xmark & \xmark & \xmark & \cmark \\ 
         InverseFinance\_2 & ETH & 02-Apr-22 & 15.6M & \xmark & \xmark & \xmark & \xmark \\ 
    \bottomrule
    \end{tabular}
    }
    \end{minipage} \hfill
    \begin{minipage}{0.48\linewidth}
    \resizebox{\textwidth}{!}{
    \begin{tabular}{llllllll}
     \toprule
        Protocol & Chain & Date & Loss & DS & DT & DF & DR \\ \midrule
        LaunchZone & BSC & 27-Feb-23 & 320K & \cmark & \xmark & \cmark & \xmark \\ 
        LUSD & BSC & 07-Jul-23 & 9K & \cmark & \xmark & \cmark & \xmark \\ 
        LW\_1 & BSC & 12-May-23 & 50K & \cmark & \xmark & \xmark & \cmark \\ 
        LW\_2 & BSC & 12-May-23 & 48K & \cmark & \xmark & \xmark & \cmark \\ 
        Mars & BSC & 16-Apr-24 & 100K & \cmark & \xmark & \xmark & \xmark \\ 
        MBC & BSC & 29-Nov-22 & 6K & \cmark & \cmark & \cmark & \xmark \\ 
        MerlinLab & BSC & 29-Jun-21 & 628K & \cmark & \xmark & \xmark & \xmark \\ 
        MonoXFinance & ETH & 30-Nov-21 & 31M & \cmark & \xmark & \xmark & \cmark \\ 
        MRGtoken & ETH & 08-Nov-23 & 12K & \cmark & \xmark & \xmark & \cmark \\ 
        NeverFall & BSC & 02-May-23 & 74K & \cmark & \cmark & \xmark & \xmark \\ 
        Nmbplatform & BSC & 14-Dec-22 & 76K & \textcircled{2} & \cmark & \cmark & \cmark \\ 
        NOVO\_1 & BSC & 29-May-22 & 76K & \cmark & \xmark & \xmark & \cmark \\ 
        NOVO\_2 & BSC & 29-May-22 & 65K & \cmark & \xmark & \cmark & \cmark \\ 
        PancakeBunny & BSC & 19-May-21 & 45M & \cmark & \cmark & \cmark & \xmark \\ 
        PancakeHunny & BSC & 20-Oct-21 & 1.93M & \textcircled{2} & \xmark & \xmark & \xmark \\ 
        PLPManager & BSC & 24-Jul-23 & 900K & \cmark & \cmark & \xmark & \xmark \\ 
        PLTD & BSC & 17-Oct-22 & 24K & \cmark & \xmark & \cmark & \cmark \\ 
        RoeFinance & ETH & 11-Jan-23 & 80K & \cmark & \xmark & \xmark & \xmark \\ 
        SanshuInu & ETH & 20-Jul-21 & 111K & \xmark & \cmark & \cmark & \xmark \\ 
        SATX & BSC & 16-Apr-24 & 29K & \cmark & \xmark & \xmark & \cmark \\ 
        SellToken & BSC & 11-Jun-23 & 100K & \cmark & \xmark & \xmark & \cmark \\ 
        SpaceGodzilla & BSC & 13-Jul-22 & 25K & \cmark & \xmark & \cmark & \cmark \\ 
        SpartanProtocol & BSC & 01-May-21 & 30M & \cmark & \xmark & \xmark & \cmark \\ 
        Starlink & BSC & 16-Feb-23 & 12K & \cmark & \xmark & \cmark & \cmark \\ 
        StarWallets & BSC & 17-Apr-24 & 33K & \xmark & \xmark & \xmark & \xmark \\ 
        STM & BSC & 06-Jun-24 & 14K & \cmark & \cmark & \xmark & \xmark \\ 
        SturdyFinance & ETH & 12-Jun-23 & 800K & \cmark & \xmark & \xmark & \xmark \\ 
        SVT & BSC & 26-Aug-23 & 400K & \cmark & \cmark & \xmark & \cmark \\ 
        SwapX & BSC & 27-Feb-23 & 1M & \xmark & \xmark & \cmark & \xmark \\ 
        TIFIToken & BSC & 10-Dec-22 & 51K & \textcircled{1} & \cmark & \cmark & \xmark \\ 
        UN & BSC & 06-Jun-23 & 26K & \cmark & \xmark & \xmark & \cmark \\ 
        UPSToken & ETH & 18-Jan-23 & 45K & \cmark & \cmark & \cmark & \cmark \\ 
        Upswing & ETH & 17-Jan-23 & 36K & \cmark & \cmark & \cmark & \xmark \\ 
        uwerx\_network & ETH & 02-Aug-23 & 324K & \cmark & \xmark & \xmark & \cmark \\ 
        UwULend & ETH & 10-Jun-24 & 19M & \cmark & \xmark & \xmark & \xmark \\ 
        ValueDeFi & ETH & 14-Nov-20 & 6M & \cmark & \cmark & \cmark & \cmark \\ 
        VesperFinance & ETH & 02-Nov-21 & 2M & \xmark & \xmark & \xmark & \xmark \\ 
        VINU & ETH & 06-Jun-23 & 6K & \cmark & \xmark & \cmark & \xmark \\ 
        WarpFinance & ETH & 17-Dec-20 & 7.8M & \cmark & \cmark & \xmark & \cmark \\ 
        WGPT & BSC & 12-Jul-23 & 80K & \cmark & \xmark & \cmark & \cmark \\ 
        WienerDoge & BSC & 25-Apr-22 & 30K & \cmark & \cmark & \cmark & \cmark \\ 
        XSTABLE & ETH & 09-Aug-22 & 56K & \cmark & \xmark & \cmark & \xmark \\ 
        Z123 & BSC & 22-Apr-24 & 136K & \cmark & \xmark & \xmark & \cmark \\ 
        Zoompro & BSC & 05-Sep-22 & 61K & \xmark & \xmark & \cmark & \xmark \\ 
        ZS & BSC & 08-Oct-23 & 14K & \cmark & \xmark & \cmark & \cmark \\ 
        Zunami & ETH & 13-Aug-23 & 2M & \cmark & \xmark & \xmark & \xmark \\ 
    \bottomrule
    \end{tabular}
    }
    \end{minipage}
    \label{tab:RQ1_res}
\end{table*}

\subsection{RQ1: Detection Effectiveness}
\label{sec:RQ1}

To answer RQ1, we evaluate \tool and compare it with three SOTA tools, DeFiTainter~\cite{DeFiTainter23}, DeFort~\cite{DeFort24}, and DeFiRanger~\cite{DeFiRanger23},
\fixme{specialized in detecting price manipulations}
using the dataset \textit{D1}.
Because DeFiRanger is not open-source, we use the results reported in their paper for the attacks they evaluated; for other attacks, we re-implemented their approach based on the description~\cite{DeFiRanger23} and conducted our evaluation.
Although DeFort is also not open-source, we obtained a copy of code from its authors to run experiments.
\fixme{FlashSyn~\cite{chen2024flashsyn} and SMARTCAT~\cite{bosi2025following} were excluded as baselines because FlashSyn targets broader attacks beyond price manipulations and lacks guidelines for detecting new attacks, while SMARTCAT was unpublished and not open-source at the time of evaluation.}

\mytab\ref{tab:RQ1_res} presents the detection results. 
The first four columns list the name of the victim protocol, chain deployed, hack date, and resulting loss, respectively. 
\textit{DS}, \textit{DT}, \textit{DF} and \textit{DR} denote \tool, DeFiTainter, DeFort and DeFiRanger, respectively.
Note that one protocol may face multiple attacks, so we add numerical suffixes to protocol names to differentiate them. We use \cmark to indicate an attack can be successfully detected by a tool, and \xmark to indicate a detection failure.

\tool can detect most of the price manipulation attacks.
It achieves a recall rate of 80\%, outperforming all other tools. 
Overall, \tool detected 76 attacks, followed by DeFort with 50 and DeFiRanger with 49 attacks, respectively, while DeFiTainter detected only 34 attacks.
\fixme{In total, \tool failed to detect 19 cases. 
We manually investigated their root cause and found: (a) 11 cases should be detectable but were missed by \tool due to the limitation of LLM reasoning, non-ERC20 token and the restriction to only single-transaction analysis, and (b) 8 cases cannot be detected because they do not satisfy the source code requirement of \name due to either \textcircled{1} missing source code or \textcircled{2} compilation errors, as marked in \mytab\ref{tab:RQ1_res}.}
\fixme{Moreover, the statistical significance of \tool's performance improvements over each baseline is confirmed by the McNemar Test~\cite{mcnemar1947note}, with all the resulting $p$-values being below 0.05 (see details in ~\cite{supplementary_material}).}
\myfig\ref{fig:price_model_proportion} details the performance of each tool on the evaluated DeFi protocols across four application categories using different price models.
Compared to other tools, \tool performs the best in every application category.
Particularly, it achieves the highest recall, 90.7\%, in 
the Token category.
However, \tool yields a low recall rate in Lending-related protocols, though still higher than all existing tools. 
Through further analysis, we identifies that 3 out of 12 attacks (InverseFinance\_2, SanshuInu, and VesperFinance) are cross-transaction attacks that \tool is unable to detect.
Additionally, one attack (TIFIToken) involves exploiting a closed-source custom price model, and the InverseFinance\_1 attack, a false negative, will be detailed below.
\fixme{After analyzing 11 detectable but missed attacks,}
we discovered that 8 out of 11 are cross-transaction attacks~\cite{ATK,Inuko,ARK,Inverse,SanshuInu,StarWallets,SwapX,Vesperfinance}, where detection was unsuccessful because \tool is based on analyzing individual transactions. For Zoompro, detection failure occurred because the token involved in the transaction did not adhere to the ERC20 token standard, resulting in the transfer event not being identified. The remaining two attacks, i.e., ``IndexedFinance'' and ``InverseFinance\_1'' cases, were subject to in-depth analysis.
The detection incapability for IndexedFinance was due to its use of an extremely complicated pricing mechanism that involves exponential calculations in its price-related function \texttt{joinswapExternAmountIn}. In the case of InverseFinance, the attacker exploited the flawed price dependency when calculating the price of tokens deposited as collateral, where the collateral price is based on the balance of multiple tokens in the liquidity pool and off-chain price oracles. As both cases require precise quantitative calculations, \tool is limited by the current LLMs' constrained capacity for scientific computation. A potential solution could be to integrate with Program-Aided Language models (PAL)~\cite{gao2023pal}, guiding the LLM to generate scripts for necessary calculations and executing them to obtain the result.


\fixme{Among the 8 undetectable transactions,}
3 were not analyzed successfully due to the unavailability of the code of price calculation functions since our method relies on code-level analysis; 
five cases involved compilation errors during the extraction of price calculation functions using Slither~\cite{slither}, an off-the-shelf static analyzer for Solidity and Vyper, thereby prematurely terminating the detection process. 
These limitations are not inherent to \tool's methodology and can be mitigated by automated or semi-automated techniques, e.g., code decompilation~\cite{grech2019gigahorse,grech2022elipmoc,suiche2017porosity} and manual intervention.


\begin{figure}[t]
    \centering
    \small
    \includegraphics[trim={0 3pt 0 3pt}, clip, width=0.8\linewidth]{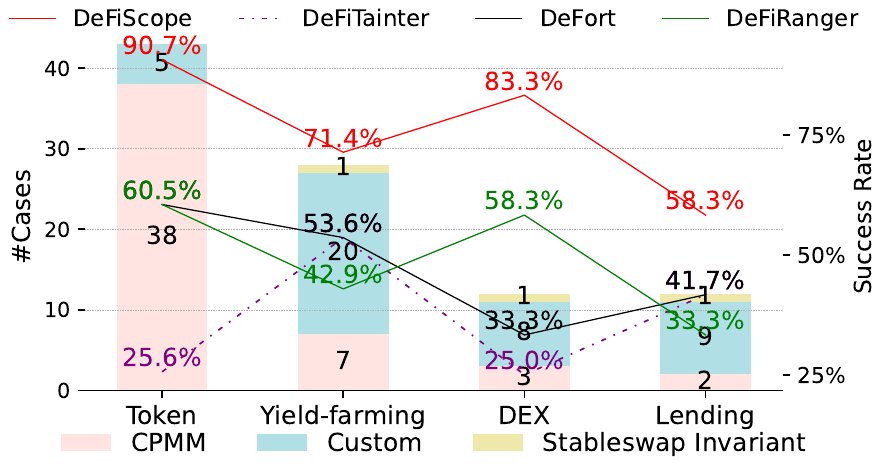}
    \vspace{-2ex}
    \caption{The categorized detection results on DeFi protocols.}
    \label{fig:price_model_proportion}
\end{figure}

\subsection{RQ2: Ablation Study}
\label{sec:RQ2}

\begin{figure}[t]
\centering
    \vspace{-2ex}
    \subfloat[GPT-3.5-Turbo w/o fine-tuning]{
        \includegraphics[width=.42\columnwidth]{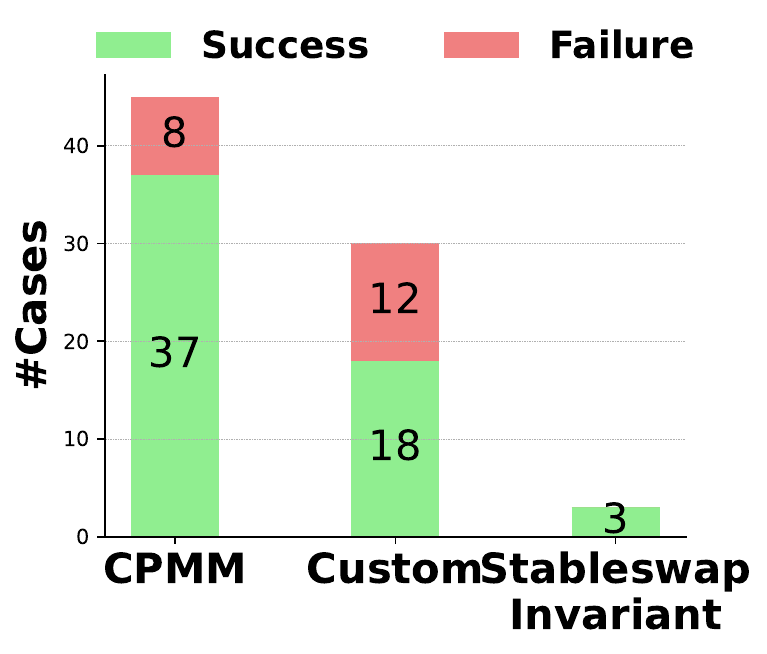}
    }
    \subfloat[GPT-3.5-Turbo w/ fine-tuning]{
        \includegraphics[width=.42\columnwidth]{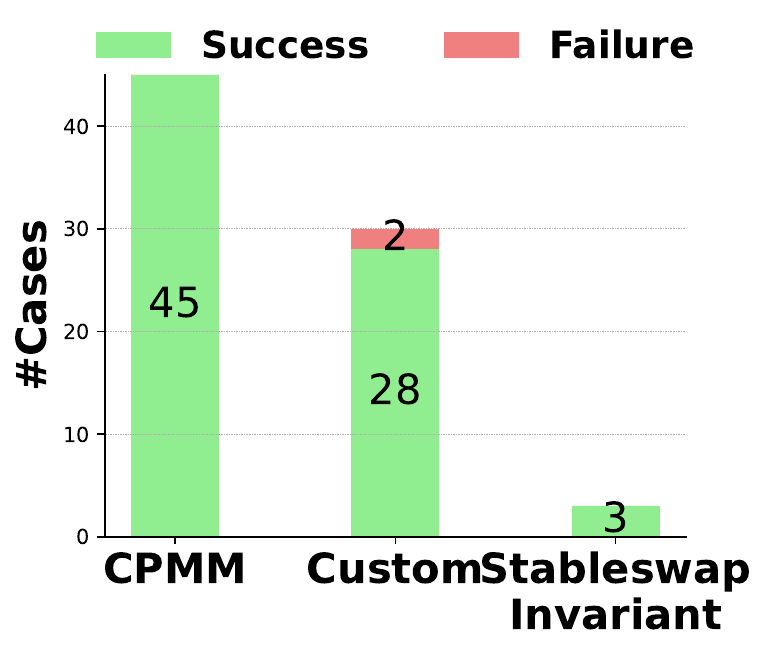}
    }
    \vspace{-2ex}
    \qquad
    \subfloat[GPT-4o w/o fine-tuning]{
        \includegraphics[width=.42\columnwidth]{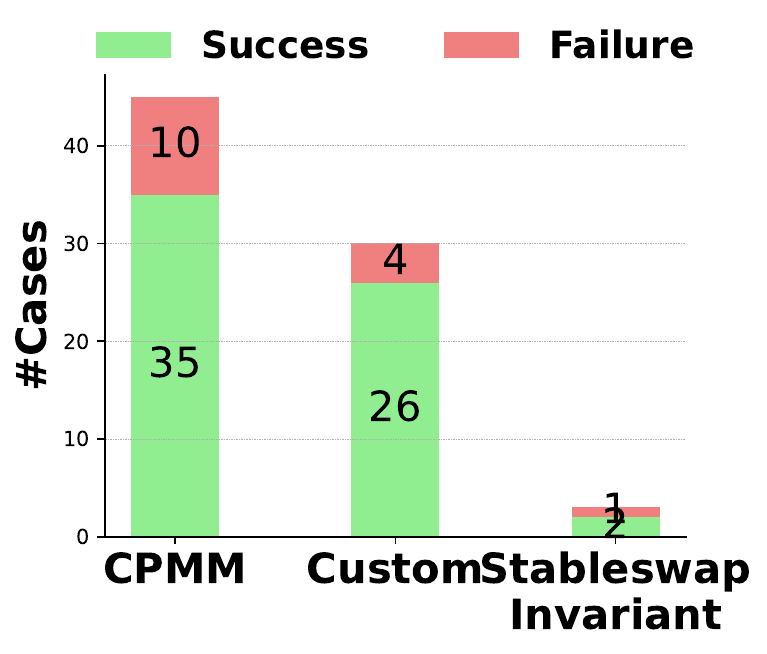}
    }
    \subfloat[GPT-4o w/ fine-tuning]{
        \includegraphics[width=.42\columnwidth]{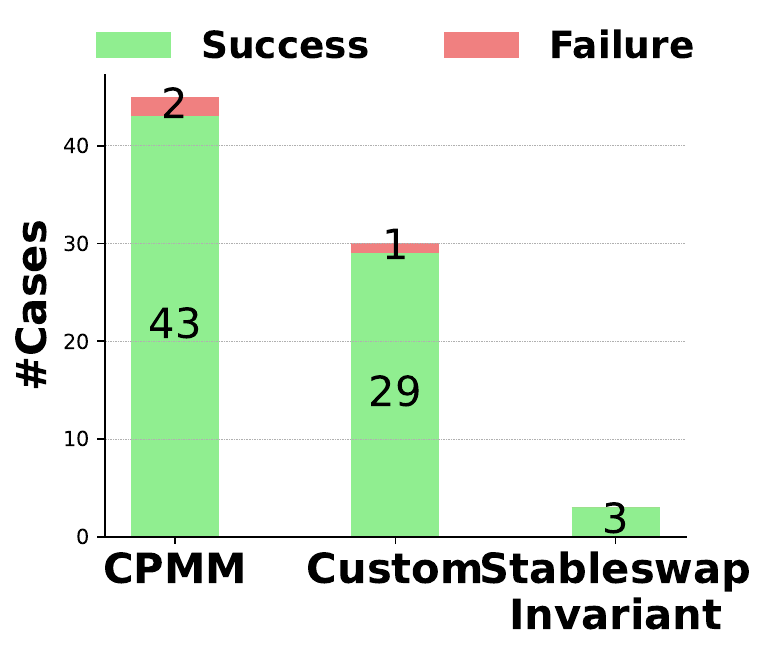}
    }
    \caption{Effectiveness of fine-tuning for 78 attacks that go through \name's LLM inference. Cases not plotted in the figures include eight due to missing source code and compilation errors, and nine due to cross-transaction issues and non-adherence to the ERC20 token standard, as mentioned in \mysec\ref{sec:RQ1}.}
    \label{fig:RQ_2}
\end{figure}

In this RQ, we investigate how fine-tuning can enhance \name's detection accuracy on the same \textit{D1} dataset, as well as the impact and cost of fine-tuning different GPT models.
To this end, we test four settings shown in \myfig~\ref{fig:RQ_2}: (a) the original GPT-3.5-Turbo without fine-tuning, (b) GPT-3.5-Turbo with fine-tuning, used in RQ1, (c) the original GPT-4o without fine-tuning, and (d) GPT-4o with fine-tuning.

\myfig~\ref{fig:RQ_2} shows that fine-tuning significantly enhances \name's ability to detect attacks, with the fine-tuned versions of GPT-3.5-Turbo and GPT-4o detecting 18 (31\%) and 12 (19\%) more attacks, respectively.
It also indicates that fine-tuning provides a more noticeable improvement for less powerful models, such as GPT-3.5-Turbo, compared to stronger models like GPT-4o.
With fine-tuning, the detection success rate for attacks targeting CPMM increases to 100\% with GPT-3.5-Turbo and 95.6\% with GPT-4o\footnote{GPT-4o does not exhibit an advantage with CPMM but shows a clear advantage with custom price models, especially for the original GPT-4o.}. 
This very high rate can be attributed to the use of CPMM during the fine-tuning training phase, enabling the model to effectively handle it. 

Although \name's fine-tuned models were trained using only CPMM data, we find that they also exhibit strong transfer learning capabilities for attacks targeting custom price models.
Specifically, the detection success rate for attacks targeting custom price models increases from 60\% to 93.3\% (a relative increase of 55.5\%) for GPT-3.5-Turbo and from 86.7\% to 96.7\% (a relative increase of 11.5\%) for GPT-4o. 
Further analysis of the LLM responses generated during detection reveals that the most significant difference introduced by fine-tuning is that the fine-tuned model strictly adheres to the CoT approach we specified.
This involves initially extracting the price model from the given code and then conducting inference based on the extracted model along with the provided information about balance changes.
In contrast, although the original model produces the final evaluation scores, it does not strictly follow the instructions of the CoT prompts.


\noindent
\textbf{LLM Costs.}
\mytab\ref{tab:compare_between_models} highlights that while the original GPT-4o generally performs better than GPT-3.5-Turbo, this performance gap can be largely minimized through fine-tuning.
However, regarding costs, GPT-3.5-Turbo has a clear advantage, which is why \name uses GPT-3.5-Turbo for fine-tuning by default instead of GPT-4o.
Specifically, the fine-tuning cost of GPT-3.5-Turbo is 68\% lower than that of GPT-4o, and the per-request inference cost of GPT-3.5-Turbo is merely \$0.0107~\cite{OpenAI_Price}, which is also 18\% cheaper than that of GPT-4o.
Therefore, for a tradeoff between cost and performance, we recommend using GPT-3.5-Turbo. 

\begin{table}[t!]
    \centering
    \caption{Comparison between GPT-3.5-Turbo and GPT-4o.}
    \vspace{-2ex}
    \small
    \resizebox{\columnwidth}{!}{
    \begin{tabular}{lcccc}
    \toprule
         Model & $TP$ & $Recall$ & \makecell[c]{Ave. Cost(\$)/\\per inference} & \makecell[c]{Fine-tuning\\Cost(\$)} \\
        \midrule
        GPT-3.5 w/o fine-tuning & 58 & 0.61 & 0.0023 & -\\
        GPT-3.5 w/ fine-tuning & 76 & 0.80 & 0.0107 & 8\\
        \midrule
        GPT-4o w/o fine-tuning & 63 & 0.66 & 0.008 & -\\
        GPT-4o w/ fine-tuning & 75 & 0.79 & 0.0131 & 25\\
    \bottomrule
    \end{tabular}
    }
    \label{tab:compare_between_models}
\end{table}

\subsection{RQ3: Real-World Practicality}
\label{sec:RQ3}


To be practical in a real-world setting, \name not only needs to maintain high detection rates for true attacks but also minimize false alarms for benign transactions.
In this RQ, we evaluate this aspect of \name and the associated time overhead using the datasets \textit{D2} and \textit{D3}, which were introduced in the prologue of \mysec\ref{sec:evaluation}.
Specifically, \textit{D2} comprises 968 suspicious transactions with various attacks (i.e., not limited to price manipulations), and \textit{D3} includes 96,800 benign transactions collected from DeFort~\cite{DeFort24}'s dataset. 

For \textit{D2}, \tool flagged 153 price manipulation attacks out of 968 suspicious transactions.
To robustly confirm these potential attacks, we cross-referenced them with attack reports or alerts published by security companies through their official channels~\cite{ancilia_twi,beosin_twi,blocksec_twi,slowMist_twi}, thereby verifying the root causes of the attacks.
Using this method, we confirm that 66 of them are previously reported price manipulation attacks.
For the remaining cases, we conducted comprehensive and in-depth analysis by combining manual review with ancillary evidence, such as identifying whether the EOA initiating the transaction was marked as a hacker by blockchain explorers~\cite{bscscan,etherscan,oklink}.
\fixme{For newly discovered incidents, we manually verified: (i) whether abnormal token price or exchange rate changes occured, and (ii) whether the user gained profit from such changes. Cases satisfying both criteria were labeled as price manipulation attacks.}
Finally, we discovered 81 previously unknown historical incidents, which were neither reported by security companies nor tagged as malicious transactions by blockchain explorers, and identified a total of six false positive cases 
(the full list can be found in~\cite{supplementary_material}).
Of these, five are benign transactions initiated by the same EOA and exhibiting identical invocation flows, while the last one is indeed an attack due to a logic issue rather than a price manipulation attack, resulting in a precision of 96\% (147/153)
\fixme{on suspicious transactions.}
In comparison, DeFort identified only 58 attacks, of which 9 were false positives labeled as other vulnerabilities by security researchers.
Moreover, DeFort failed to detect 114 of the malicious transactions flagged by \tool, including 54 that are officially confirmed attacks.

We further analyze \name's false alarms in a realistic setting mixing 968 suspicious transactions in \textit{D2} with 96,800 benign transactions in \textit{D3} (see the earlier dataset setup in the prologue of this section).
For each transaction, we set the maximum scan time to 300 seconds.
The results reveal that \tool achieves zero false alarms on benign transactions in \textit{D3},
\fixme{and 6 false positives on \textit{D2} (as mentioned above),}
with an average of 2.5 seconds per transaction across all transactions. 
\fixme{More than half the time is spent on price change inference that uses up an average of 1.40 seconds, while static analysis takes slightly less, around 0.97 seconds on average. The other steps within \name incur negligible cost. A detailed time cost breakdown is available in ~\cite{supplementary_material}.}
This highlights \name's potential for large-scale, daily on-chain monitoring scenarios.
\section{Discussion}
\label{sec:discussion}


\noindent\textbf{Cross-LLM Generalizability.}
To show \name's methodology is not limited to OpenAI's fine-tuning paradigm~\cite{OpenAI_Fine-tuning}, we performed SFT fine-tuning~\cite{chung2024scaling} on a open-source
Phi-3~\cite{hf_phi_3_128k_instruct}
model using LoRA~\cite{hu2021lora},
on a sever
with 4 Nvidia H800 GPUs.
We used \textit{all}
1,000 price change
samples mentioned in \mysec\ref{sec:priceInfer},
split 8:1:1 into training, evaluation, and test sets.
As shown in \mytab\ref{tab:comparison_btw_phi_and_gpt},
LoRA-based fine-tuning enabled Phi-3 to
outperform
GPT-3.5-Turbo (without fine-tuning) and
achieve performance comparable
to GPT-4o (without fine-tuning).
Full hyperparameter settings and results are available in our repository~\cite{defiscope_model_comparison_result}.

\begin{table}[t]
    \centering
    \small
    \caption{Comparison between GPT tuning and Phi-3's SFT.}
    \vspace{-2ex}
    \resizebox{0.49\textwidth}{!}{
        \begin{tabular}{lclc}
        \toprule
            Model  &  $Recall$ & Model  &  $Recall$\\
            \midrule
            GPT-3.5 w/o fine-tuning & 0.61 & GPT-3.5 w/ fine-tuning & 0.80 \\
            GPT-4o w/o fine-tuning & 0.66 & GPT-4o w/ fine-tuning & 0.79 \\
            Phi-3 w/o fine-tuning & 0.52 & Phi-3 w/ fine-tuning\fixme{~\cite{ft_phi_3_hg}} & 0.66 \\
            \bottomrule
        \end{tabular}
    }
    \label{tab:comparison_btw_phi_and_gpt}
\end{table}

\fixme{
\noindent
\textbf{Cross-Blockchain Generalizability.}
Although our evaluation primarily focused on Ethereum and BSC, which together account for the majority of observed price manipulation attacks, \name is generalizable to EVM-compatible blockchains, such as Polygon~\cite{polygon} and Arbitrum~\cite{arbitrum}. \name successfully detected seven price manipulation attacks
on these chains. Full details of these cases are available in~\cite{supplementary_material}.
}

\noindent\textbf{\fixme{Closed-source contracts and cross transaction attacks.}}
The availability of price model code can affect \tool's detection accuracy.
According to our study, most DeFi applications are open-source to gain user trust.
To mitigate issues with closed-source price models, we design the Type-II prompt to cover price models in those closed-source liquidity pools.
\fixme{Currently, \tool detects only single-transaction attacks, while some attacks span multiple transactions to bypass protocol time restrictions~\cite{chen2024demystifying}.
Detecting such cases is challenging. For instance, attacks like \cite{Inuko} span over 48 hours (about 57,000 blocks), making it difficult to trace all related transactions. Fortunately, our study observed that such attacks are far less common than single-transaction ones.}
\section{Related Work}
\label{sec:related}

\noindent
\textbf{On-chain Security Analysis.}
Prior efforts~\cite{zhou2021high, qin2021attacking, qin2022quantifying, daian2020flash, zhou2023sok} have studied blockchain threats and DeFi attacks.
Some transaction-based systems are proposed to mine vulnerable transaction sequences~\cite{zhang2020txspector, so2021smartest}, explore arbitrage opportunities~\cite{zhou2021just}, detect malicious phishing~\cite{he2023txphishscope}, and simulate attacks to prevent intrusions~\cite{qin2023blockchain}.
Tools like FlashSyn~\cite{chen2024flashsyn}, DeFiRanger~\cite{DeFiRanger23}, DeFiTainter~\cite{DeFiTainter23}, and DeFort~\cite{DeFort24} are capable of detecting attacks associated with price manipulation.
FlashSyn uses numerical approximation to synthesize malicious contracts that target DeFi apps through price manipulation attacks.

\noindent
\textbf{LLMs for Smart Contract Security.}
LLMs have become powerful tools in blockchain security.
GPTScan~\cite{sun2024gptscan} leverages LLMs with static analysis to detect logical vulnerabilities.
BlockGPT~\cite{gai2023blockchain} uses LLMs to rank transaction anomalies for real-time intrusion detection.
LLM4Vuln~\cite{sun2024llm4vuln} improves LLM reasoning for vulnerability analysis.
iAudit~\cite{ma2024combining} combines fine-tuning with agents for intuitive auditing.
PropertyGPT~\cite{liu2024propertygpt} uses LLMs for retrieval-augmented property generation.
To our knowledge, \tool is the first tool that uses LLMs specifically designed for detecting price manipulation attacks.

\section{Conclusion}
\label{sec:conclusion}

In this paper, we introduced \tool, the first tool that utilizes LLMs specifically for detecting price manipulation attacks.
\tool employs LLMs to intelligently infer the trend of token price changes based on balance information within transaction executions.
To strengthen LLMs in this aspect, we simulated on-chain transaction data and fine-tuned a DeFi price-specific LLM.
We also proposed a graph-based method to recover high-level DeFi operations and systematically mined eight price manipulation patterns.
Our evaluation demonstrated \tool's superior performance over SOTA approaches and real-world impact.
Future work includes better handling of closed-source price calculation functions.

\ifCLASSOPTIONcompsoc
\section*{Acknowledgments}
\else
\section*{Acknowledgment}
\fi
\addcontentsline{toc}{section}{Acknowledgment}
We thank all reviewers for their constructive feedback.
This research was supported by Lingnan Grant SUG-002/2526, HKUST TLIP Grant FF612, the National Natural Science Foundation of China (Project No. 72304232), the Singapore Ministry of Education Academic Research Fund Tier 2 (T2EP20224-0003) and the Nanyang Technological University Centre for Computational Technologies in Finance (NTU-CCTF). Any opinions, findings, and conclusions or recommendations expressed in this material are those of the author(s) and do not necessarily reflect the views of MOE and NTU-CCTF.



%
\bibliographystyle{IEEEtran}   
\bibliography{main}


\end{document}